# Crude Oil Displacement Enhanced by Interfacially Active Nanoparticles and Their Coupling Effect with Low-Salinity Brines

Suparit Tangparitkul,[1,*] Anupong Sukee,[2] Jiatong Jiang,[3,4] and David Harbottle[5]

[1]Department of Mining and Petroleum Engineering, Faculty of Engineering, Chiang Mai University, Chiang Mai 50200, Thailand
[2]Department of Alternative Energy Development and Efficiency, Ministry of Energy, Bangkok 10330, Thailand
[3]School of Petroleum, China University of Petroleum (Beijing) at Karamay, Karamay 834000, China
[4]Unconventional Petroleum Research Institute, China University of Petroleum (Beijing), Beijing 102249, China
[5]School of Chemical and Process Engineering, University of Leeds, Leeds, LS2 9JT, UK

*To whom correspondence should be addressed: Email suparit.t@cmu.ac.th Phone: +66 5394 4128 Ext. 119

**Abstract**

From the microscopic scale to the petroleum-reservoir scale, the interfacial phenomena of the crude oil-water-rock system crucially control an immiscible flow in a porous reservoir. One of the key mechanisms is crude oil droplet displacement dynamics, which can be optimized by manipulating the oil-water interfacial tension and the three-phase contact angle by means of chemical injection. The current study primarily investigated oil displacement enhanced by interfacially active nanoparticles, namely poly(*N*-isopropylacrylamide) or pNIPAM (43 - 48 nm hydrodynamic diameter and 0.0005 - 0.0050 wt.% concentration range), which found an acceleration of oil droplet receding rate (5.66 °/s) and a greater degree of oil droplet dewetted (37.0° contact angle). This was due to a contribution from the nanoparticle-induced structural disjoining pressures between the oil-water and water-solid interfaces. The coupling effect of pNIPAM nanoparticles with low-salinity brines was examined, which revealed a discrepancy in different brine valences. Coupling with divalent $CaCl_2$ led to much slower oil droplet receding dynamics (2.58 °/s, and 58.7° contact angle) since oil-substrate bridging is formulated and promoted by the divalent cation. However, positive synergy was observed with a monovalent NaCl blend. The crude oil dewetting dynamics were enhanced (9.55 °/s) owing to the combined salt-induced hydration and nanoparticle-induced structural forces. The contact angle was as low as 21.5° before eventually detaching from the substrate for a relatively short period (156 s). These findings highlight the coupling effect of nanoparticles and low-salinity brine on the dewetting of heavy crude oil. Adding nanoparticles to an 'optimal' brine could be an option for faster and greater fluid displacement, which is not limited to oil production applications but several others, such as detergency and other forms of geological storage.

# 1. Introduction

Transitioning away from fossil fuels implies urgent petroleum consumption to get to the peak as quickly as possible [1]. Such a global consensus acknowledges the need for affordable energy, especially for developing economies, with concurrent ambition toward cleaner energy alternatives. Thus, exploiting existing petroleum reserves is an inevitable approach to ensure fairness in this energy transition, with many petroleum producers implementing the tertiary stage of enhanced oil recovery (EOR).

The prime mechanism of EOR is the enhancement of microscopic oil displacement, which is based on fundamentals across scale, including interfacial forces, microscopic interactions of the solid-oil-water system, and fluid dynamics in porous media. Of those fundamentals, the spontaneous dewetting of crude oil droplets on solid substrate has long been researched, not only to elucidate the theoretical background of its interfacial phenomena but also to serve as a study examination bridging sub-microscopic driving forces to physical microscale phenomena [2,3].

Droplet dewetting dynamics have been widely theorized into two models, namely hydrodynamic (HD) and molecular-kinetic (MK) [2,4,5], which usually describe how the non-wetting oil droplet recedes from a solid substrate as influenced by the existence of displacing wetting aqueous phase. The HD model describes a creeping flow of viscous fluid (e.g., heavy crude oil) in the vicinity of the three-phase contact line, assuming the slip boundary condition of the fluid-fluid contact line on a solid substrate. The displacement velocity of the three-phase contact line ($v$) expresses as a function of the oil-water interfacial tension ($\sigma$) and the dynamic contact angle ($\theta_d$) [4,6]:

$$v = \frac{\sigma}{9\mu_0}[(\pi - \theta)^3 - (\pi - \theta_d)^3]\left[\ln\left(\frac{L}{L_s}\right)\right]^{-1} \qquad (1)$$

where $\mu_0$ is the oil viscosity, $\theta$ the steady-state contact angle measured through the water phase [2], and $L_s$ and $L$ the characteristic length of the oil droplet and effective slip length, respectively. The MK considers molecular displacements of adsorption and desorption in the vicinity of the three-phase contact line, of which fluid molecules can detach and attach to neighboring sites if an energy barrier is overcome. The relationship between the $\theta_d$ and the $v$ based on the MK mechanism can be simplified, as follows [5,7]:

$$v = \frac{\sigma}{\zeta}[\cos(\pi - \theta_d) - \cos(\pi - \theta)] \qquad (2)$$

where $\zeta$ is the coefficient of the contact-line friction, treated as an adjustable parameter of experimental data.

Previous research has shown that the interfacial phenomena (namely, $\sigma$ and $\theta$) could dramatically influence the droplet dynamics, as described by the theoretical models above [4,5]. This emphasizes that an approach to "engineer" crude oil droplet displacement can be improved by modifying such oil-water-rock interfacial properties. Recent studies have demonstrated how the contributions from brine valency and concentration control such droplet dynamics of heavy crude oil [8–10], highlighting insights into "low-salinity brine" or "ion-tuned" water EOR. Manipulated brines were found to alter the $\sigma$ and $\theta$ substantially, although optimal brine components to yield effective alterations are system-specific (e.g., different from crude oil to crude oil) [11–13]. A greater reduction in $\sigma$ was observed with divalent brines (e.g., $CaCl_2$) rather than monovalent (e.g., NaCl). However, monovalent brines likely altered the $\theta$ to a higher degree as a contribution from a non-DLVO hydration force, and screened the influence of the $\sigma$ reduction on the droplet dewetting dynamics [9,10,14]. The study concluded that the $\theta$ alteration governs the crude oil droplet dewetting dynamics in the studied system of brines, with the theoretical modeling emphasized [10]. To reduce the $\sigma$ dramatically, interfacially active agents (e.g., surfactants) are added to aqueous phase in the EOR process. Microscopically, the reduced $\sigma$ dictates the oil droplet dynamics, as described in the equations above [2,15].

In addition to the engineered brines and surfactant fluids, much research has also investigated the potential of nanofluids that could contribute to changes in both $\sigma$ and $\theta$ [16–18], despite several controversial observations reported [19,20]. With nanoparticle-surface functionalization, the nanoparticles dispersed in the bulk fluid are reasoned to achieve an ability to partition at the fluid-fluid interface and lead to a reduction in $\sigma$ [17,21], though this $\sigma$ reduction as of energetic-favorably is questioned [19,22,23]. Another well-documented mechanism is a nanoparticle-induced structural disjoining pressure, which repels the oil-water interface from the oil-substrate interface effectively at the water wedges. Thus, oil displacement improved with apparent $\theta$ decreased [16,18].

Given the energetic-favorably interfacially active property, the nanogel compound, namely poly(*N*-isopropylacrylamide) or pNIPAM nanoparticles, were examined as potential system for EOR [24–28]. Many investigations have been carried out on pNIPAM interfacial interactions at the oil-water interface, though the interactions at the three-phase contact line (e.g., structural disjoining pressures and $\theta$) should be explored further. In both model oil and heavy crude oil systems, pNIPAM nanoparticles demonstrated a great adsorption ability at the oil-water interface, leading to a significant reduction in $\sigma$ [25,26,28,29]. In previous work by the authors, the competitive adsorption of pNIPAM and anionic surfactant (sodium dodecyl sulfate: SDS) at the heavy crude oil-water interface was found to be a function of total pNIPAM-SDS component concentration, with each species contribution quantified [28]. At a low component concentration (0.0005 wt.%), the SDS partitions at the oil-water interface and the pNIPAM nanoparticles were likely dispersed in the bulk fluid. At higher concentrations (0.0050 wt.%), pNIPAM nanoparticles partitioned at the oil-water interface and dominantly contributed to a greater reduction in $\sigma$ when compared to that of the low concentration. Such a result implies the existence of a higher number of pNIPAM in the bulk phase at a lower concentration, with nanoparticle-induced structural disjoining pressures anticipated, thus showing improvement in oil displacement dynamics. Nevertheless, a systematic oil dewetting experiment is needed.

In the current study, the dewetting dynamics of heavy crude oil droplets were studied with the set of pNIPAM-SDS fluids investigated previously [28] to examine their ability on crude oil droplet displacement and investigate a potential contribution from the nanoparticle-induced structural disjoining pressures. Furthermore, the coupling effect of the nanofluids with low-salinity brines (2000 ppm), which is important to field applicability, will be assessed on the pNIPAM nanoparticle workability in the brines to observe any synergistic contribution to the dewetting dynamics process – the combined salt-induced and nanoparticle-induced structural forces. The latter process is the main novelty of the current work to elucidate such an interfacial phenomenon in the existence of brine.

## 2. Materials and Experimental Methods

### 2.1 Chemicals and nanoparticle synthesis

Heavy crude oil with 39.5% asphaltene was used as an oil phase throughout the current study as a dead oil phase. It is important to note that the oil phase used in the current study is the same as the previous studies [10,28], aiming for comparison benefits. The oil has a density of 946 kg/m$^3$ (18.03° API), and a viscosity of 363.7 mPa·s measured at 60 °C. Anionic SDS surfactant (Sigma-Aldrich, UK) was used as received. *N*-isopropylacrylamide (NIPAM, Sigma-Aldrich, UK), *N,N'* Methylenebisa-crylamide (BA, Fluka), and Potassium persulfate (KPS, Merck) were used as substances to synthesize pNIPAM nanoparticles, as described below. Ultrapure Milli-Q water with a resistivity of 18.2 MΩ·cm and pH of 5.5 was used in all experiments. Sodium chloride as monovalent salt (NaCl ≥ 99.5%, Sigma-Aldrich, UK) and calcium chloride dehydrate as divalent salt (CaCl$_2$, ≥ 99.5%, Sigma-Aldrich, UK) were used as received to prepare the brine solutions using the Milli-Q water.

The pNIPAM nanoparticles were synthesized following the process of Li et al [25,26]. Briefly, NIPAM (2.25 g) and BA (75 mg) were dissolved into 250 mL of Milli-Q water in a 500 mL three-necked reactor and stirred at 300 rpm. After stirring the solution for 30 min at 70 °C under nitrogen bubbling, the temperature was maintained for a further 30 min. During this time, KPS (0.25 g) dissolved in 10 mL of Milli-Q water was added dropwise to initiate polymerization. The reaction was kept at 70 °C for 3.5 h and then cooled at room temperature. After that, the solution mixture was passed through glass wool to remove any agglomerated excess, and then further purified five times by centrifugation at 10000 rpm. The stock solution of pNIPAM (prepared to 0.058 wt.%) was diluted to the desired concentration before each use.

### 2.2 Nanoparticle characterization

The hydrodynamic diameter and ζ potential of pNIPAM nanoparticles were measured using a ZetaSizer Nano ZS (Malvern Instrument, UK). The measurement was evaluated within Milli-Q water, 2000 ppm NaCl, and 2000 ppm CaCl$_2$ at an experimenting temperature of 60 °C. Twelve measurements were conducted to obtain the average result.

### 2.3 Nanofluid systems and their preparation

The PNIPAM + SDS solution was prepared by blending pNIPAM and SDS at an equivalent mass ratio of 1:1 in Milli-Q water followed by mixing for 30 min to ensure good dispersion. According to the previous study of the measured steady-state interfacial tension of the pNIPAM + SDS blends compared to their single component systems at different total component concentrations (**Fig. 1**) [28], the two concentrations (i.e., 0.0005 wt.% and 0.0050 wt.%) of the total component were chosen for investigating their performance on the crude oil droplet dewetting dynamics in the current study. The mixtures are Newtonian fluids, with measured viscosity within 0.1 mPa·s. These two concentrations were compared due to their discrepancy in the interfacial occupancy of the pNIPAM nanoparticles, as discussed above, while their contributions to the dewetting dynamics could be systematically examined. As such, the experimenting displacing fluids are pNIPAM + SDS blends, SDS solutions, and pNIPAM nanofluids at the two total component concentrations, resulting in six displacing fluids in total, as given in **Figure 1.**

To examine the pNIPAM-SDS nanofluid workability and the coupling effect with low-salinity brines, the pNIPAM + SDS blend was mixed further with the brines (i.e., 2000 ppm NaCl and 2000 ppm $CaCl_2$), which are optimal fluids taken from the previous study [10].

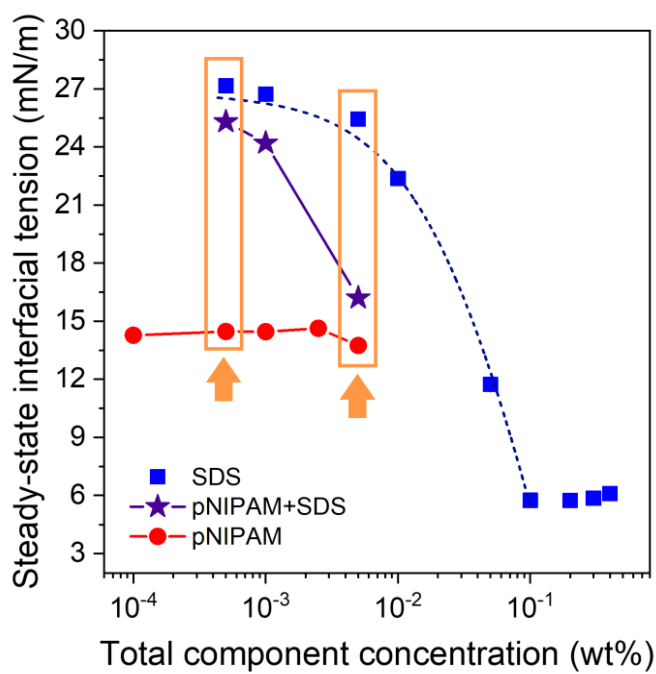

**Figure 1.** The steady-state interfacial tensions between crude oil and experimenting fluids as a function of total component concentration: pNIPAM, SDS, and pNIPAM + SDS blend. Black arrows point to the boxed fluid systems of the two concentrations (i.e., 0.0005 wt.% and 0.0050 wt.%) selected for the current study of droplet dewetting dynamics. Lines are added to guide the eye. Figure is modified from [28].

**2.4 Crude oil droplet dewetting dynamics and the steady-state contact angle**

An experiment on crude oil droplet dewetting was conducted dynamically at a controlled environment of 60 °C by means of observing the dynamic development of the droplet contact angle, following the previously established method [10]. A 10 μL sample of crude oil was deposited on a cleaned glass substrate and then placed in an insulated thermal cell. To allow only chemical additives to be examined, the model glass substrate was used in the current work to eliminate physical and chemical heterogeneities since strong influences of such heterogeneities have been found in the droplet dewetting process [30]. The glass substrate was prepared by rinsing in Milli-Q water and drying using $N_2$ gas before being placed in a UV/Ozone Procleaner™ (Bioforce Nanosciences, USA) for 15 min to remove any organics. The prepared glass had a roughness of 1.3 nm root mean square and 0.9 nm arithmetical mean deviation measured by a surface profilometer (Form Talysurf PGI 800, Taylor Hobson, UK). The experimental fluid solution was pre-heated to 60 °C and injected into the thermal cell at 1400 mL/min, being discharged onto the base of the thermal cell below the glass substrate. With a negative spreading coefficient, the oil droplet was dewetted spontaneously while the fluid was leveled up to the position of the deposited oil. The shape of the oil droplet was captured dynamically by a tensiometer camera (Theta Optical Tensiometer, Attension, Biolin Scientific, Finland) at 2 fps until the steady-state conditions were observed, where the steady-state contact angle was defined. OneAttension software was used to analyze the solid-oil-water three-phase contact angle by image analysis, and the dynamic contact angle was also reported. The reported contact angles were averaged from ≥2 repetitive experiments, and from the left and right contact angles.

## 2.5 Measurement of the crude oil-aqueous interfacial tension

The interfacial tension between the crude oil and experimental fluid was measured using the same setup as the tensiometer at 60 °C. The 10 μL of oil droplet was discharged from a stainless inverted needle (Gauge 22) using a micro-syringe pump (C201, Biolin Scientific, Finland). The oil-droplet shape was captured at 2 fps as a surface-active species partitioning at the oil-water interface until the steady state was observed (~4000 s). The interfacial tension was determined by the edge-detection method.

## 3. Results and Discussion

## 3.1 Nanoparticle characterizations

The hydrodynamic diameter of pNIPAM nanoparticles in Milli-Q water at 60 °C was measured to be ~48 nm [28]. The workability of the pNIPAM in the brines was examined against any possible aggregation. When measured in the brines, the hydrodynamic diameter decreased slightly to ~44 nm and ~43 nm for the NaCl and $CaCl_2$ systems, respectively. Such a slight shrinkage in particle size was due to the dehydration of the pNIPAM particles in excess salt environments [31–33], while the particle size shrinkage suggests no aggregation, as reported in previous work [31].

The ζ potential of pNIPAM nanoparticles was found to be negative, which is due to the contribution of the residual sulfate groups of the KPS initiator [26]. In Milli- Q water, the ζ potential was approximately -35 mV at 60 °C. In the brine systems (both NaCl and $CaCl_2$), the ζ potential was found to be between -10 and -12 mV. The magnitude of ζ potential was lower in the brines, correlating to a change in the particle size and ionic suspension [29,34,35]. The effect of additional anion and cation provides an increase in ionic strength, leading to the potential magnitude being reduced by nearly three-fold due to less charge reversal [35].

**3.2 Dewetting dynamics of crude oil droplet**

Comparisons of the dewetting dynamics of the crude oil droplet in the various fluid systems (i.e., pNIPAM + SDS blends, SDS solutions, and pNIPAM nanofluids) are shown in **Figure 2**, and the characterized performances are concluded in **Table 1**. In all fluids, the oil droplet receded due to a change in negative Gibbs free energy, and the contact angle decreased with time until it attained the steady state ($\theta$, at no later than 200 s).

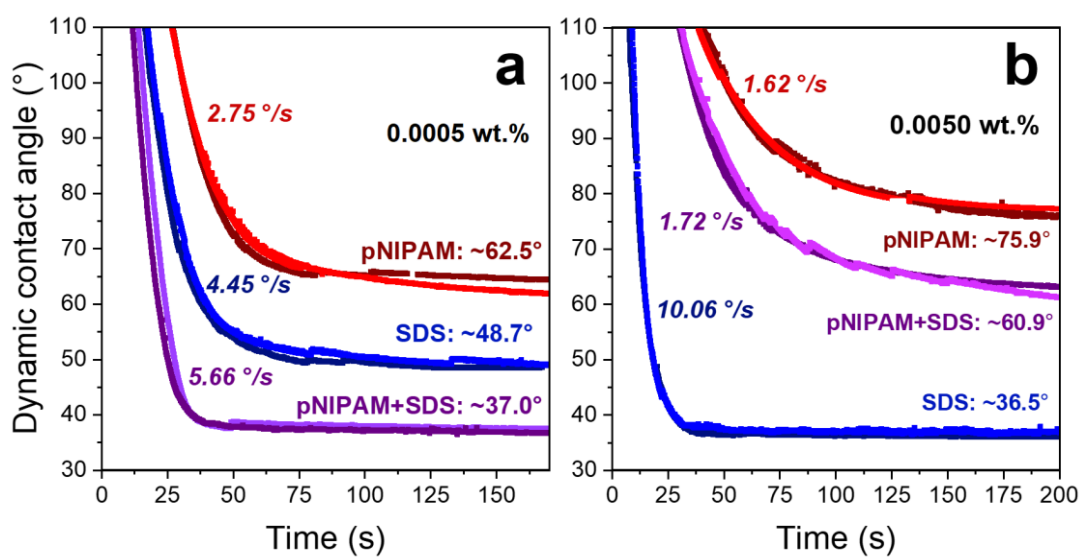

**Figure 2.** Dewetting dynamics of the crude oil droplet dewetting shown as the dynamic contact angle in the various fluid systems: pNIPAM + SDS blends (purple symbol), SDS solutions (blue symbol), and pNIPAM nanofluids (red symbol) at the total component concentration of 0.0005 wt.% (a) and 0.0050 wt.% (b). The initial receding rate and the steady-state $\theta$ are also shown with the plots.

**Table 1.** Experimental results on the dewetting performance of the oil droplet.

| Total component concentration | Component | Droplet dewetting performance | | |
|---|---|---|---|---|
| | | $\theta$ (°) | $\sigma$ (mN/m) | Initial receding rate (°/s) |
| Low concentration (0.0005 wt.%) | pNIPAM | 62.5 | 14.6 | 2.75 |
| | SDS | 48.7 | 27.2 | 4.45 |
| | pNIPAM + SDS | 37.0 | 25.3 | 5.66 |
| High concentration (0.0050 wt.%) | pNIPAM | 75.9 | 13.8 | 1.62 |
| | SDS | 36.5 | 25.4 | 10.06 |
| | pNIPAM + SDS | 60.9 | 16.2 | 1.72 |

At a low concentration (**Fig. 2a**), the receding rate of the crude oil dewetting increases and the steady-state $\theta$ decreases in the following order: pNIPAM < SDS < pNIPAM + SDS. The initial receding rate of the oil was likely not affected by the $\sigma$ since changes in the receding rate do not follow the trend in $\sigma$, as Eqs. (1) and (2) suggest, but rather are dominated by the steady-state $\theta$ (**Table 1**). In theory, $\theta$ should be decreased with declining $\sigma$, based on Young's law [36]. However, these observed results do not follow the law, implying other factors influenced the $\theta$. Such an influencing factor in the blend could potentially be a structural disjoining pressure induced by the nanoparticles [37], as the previous study confirmed that SDS is preferentially partitioned at the oil-water interface and the pNIPAM nanoparticles are mostly in the bulk fluid [28]. Given such an additional repulsive structural force, the oil droplet displacement (5.66 °/s) was accelerated when compared to the single-component systems at the same concentration and resulted in the lowest steady-state $\theta$ (37.0°). The stronger disjoining pressure between the oil-water and substrate-water interfaces secures a more stable water film, which weakens the oil-substrate affinity and 'destabilizes' or 'detaches' crude oil components from the substrate as a result [37].

The re-arrangement of pNIPAM nanoparticles at the water wedge is illustrated in **Figure 3a**, which induces the structural disjoining pressure, while the oil-water interface is secured by SDS. Further confirmation of nanoparticle accumulation near the three-phase contact line is provided by Cryo-SEM photography, as shown in **Figure 4**. The nanoparticle ordering in the water wedge of the low component system of the pNIPAM + SDS blend has been quantitatively assessed using Cryo-SEM (LEMAS, University of Leeds). The SEM image (**Fig. 4a**) showed the particle distribution from the three-phase contact line (arrowed in **Fig. 4a**) toward the water phase, indicating a uniformly dense cluster of nanoparticles near the contact line; the spread is reduced at a farther distance. The nanoparticle coverage reached about 20-25% near the contact line and decreased exponentially outward from the interface as coverage was measured at different magnifications, as reported in **Figure 4b**.

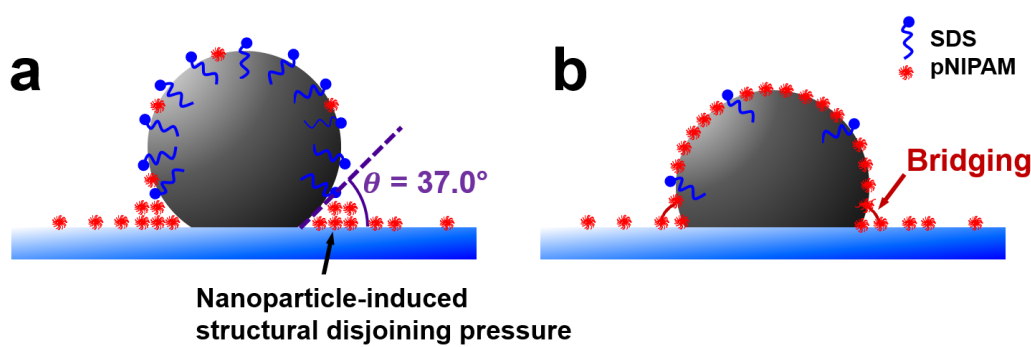

**Figure 3.** Schematic mechanisms of the crude oil droplet dewetting dynamics in the pNIPAM + SDS blends at a low component concentration (a) and high component concentration (b). Figures are not to scale.

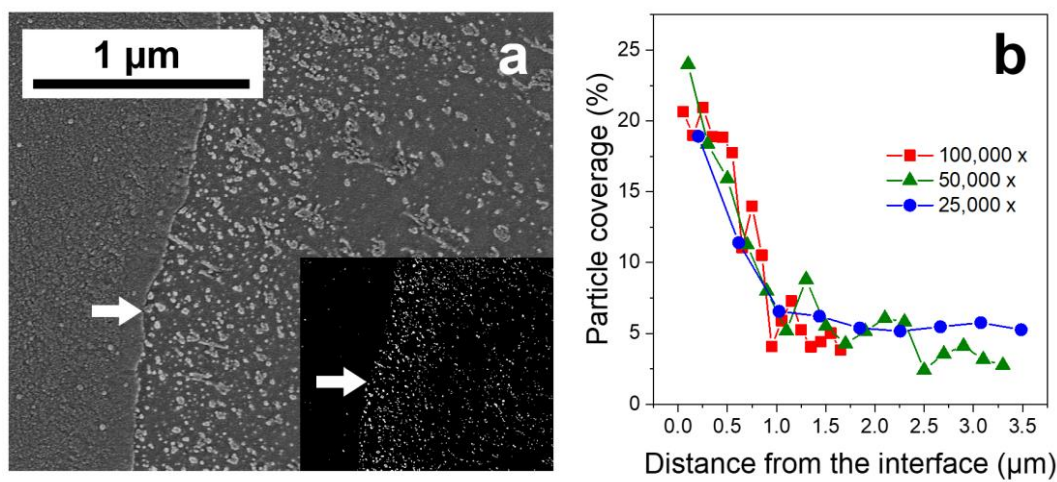

**Figure 4.** Cryo-SEM image of nanoparticle distribution from the three-phase contact line (arrowed) in the water wedge (a) with a black-and-white contrast image (inset) and the nanoparticle coverage concentration as a function of distance (b).

At a high concentration, the oil droplet receding rate increases and the steady-state $\theta$ decreases in the following order: pNIPAM < pNIPAM + SDS < SDS, see **Fig. 2b**. The SDS fluid showed a substantially high dewetting rate of 10.06 °/s, exceeding the receding rates of the pNIPAM + SDS blend and the pNIPAM-only at one order of magnitude. The oil displacement in the SDS system distinctively follows the dynamic theories described above, where high $\sigma$ and low $\theta$ promote the droplet receding dynamics – a physical contribution from a stronger repulsive force between the SDS-loaded oil-water and water-solid interfaces, while two other systems likely do so (**Table 1**).

Considering the pNIPAM + SDS blend at this high concentration, the pNIPAM was described to dominate the oil-water interface (**Fig. 1**) [28]. Such a pNIPAM-laden interface is thought to obstruct the displacement of the droplet by "sticky" polymer bridging between the oil-water and solid-water interfaces. Thus, much slower receding rates (<2 °/s) and high $\theta$ (60.9°) are observed. **Figure 3b** illustrates the droplet dynamics behavior in this high-concentration pNIPAM + SDS blend. This polymer bridging is commonly observed in the polymer-rich fluid of a crude oil-water-solid system [32]. While pNIPAM polymer stationing at the oil-water interface is concluded, the adsorption of the polymeric pNIPAM nanoparticles onto the solid substrate in the current system is confirmed by the adsorption experiment using quartz crystal microbalance, see **Supplementary Material**. A similar effect is also observed in the pNIPAM-only fluid, with more severe hindrance due to a higher amount of polymeric nanoparticles without anionic SDS, see **Fig. 2b**.

## 3.3 Dewetting dynamics performance in the low-salinity brines

Due to the secured mechanism on the structural disjoining pressure at low component concentration, the pNIPAM + SDS blend at 0.0005 wt.% was selected so its performance could be examined concurrently with the existence of the brines in this section. Thus, three displacing fluids are considered: (i) the blend in the Milli-Q water; (ii) the blend in the 2000 ppm NaCl brine; and (iii) the blend in the 2000 ppm $CaCl_2$ brine. The dewetting dynamics performances are shown in **Figure 5** and concluded in **Table 2**.

The droplet dewetting dynamics were influenced by both brines differently. In the divalent brine, the crude oil droplet receded at a slower rate and attained a higher steady-state $\theta$, even though the receding process took a longer time for the steady state. No improvement in the displacing performance was observed in any aspect when the blend dispersed in the divalent brine environment, but it was worse. Difficulty in displacing the crude oil from the substrate was likely due to the strengthened adhesion work between the crude oil and the substrate as contributed by the increased hydrophobicity of the substrate. In addition to pNIPAM adsorption, the negatively charged SDS molecules could now bind to the negatively charged substrate via the divalent cation bridging, which exists in the $CaCl_2$ brine system [32,33].

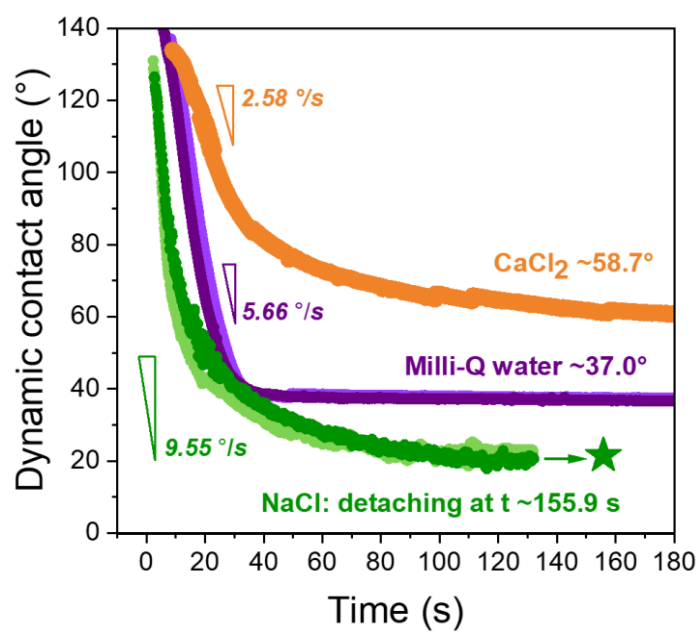

**Figure 5.** Dewetting dynamics of the crude oil droplet dewetting shown as the dynamic contact angle in the pNIPAM + SDS blends at the total component concentration of 0.0005 wt.%, dispersed in the Milli-Q water (brine-free, purple symbol), 2000 ppm NaCl brine (green symbol), and 2000 ppm $CaCl_2$ brine (orange symbol). The initial receding rate and the steady-state $\theta$ are also shown in the plots. The green star indicates the detachment of the crude oil droplet from a solid substrate in 2000 ppm NaCl brine at $t = 155.9$ s.

**Table 2.** Experimental results on the dewetting performance of the oil droplet compared in different brines.

| In solution | Droplet dewetting performance | | |
|---|---|---|---|
| | $\theta$ (°) | $\sigma$ (mN/m) | Initial receding rate (°/s) |
| Milli-Q water (brine-free) | 37.0 | 25.3 | 5.66 |
| 2000 ppm NaCl brine | detached | 12.3 | 9.55 |
| 2000 ppm CaCl$_2$ brine | 58.7 | 10.8 | 2.58 |

On the contrary, in the 2000 ppm NaCl brine, great oil droplet displacement was observed with a higher initial receding rate (9.55 °/s), and the oil droplet eventually detached from the substrate at $t \sim 155.9$ s (**Fig. 6**), with the last apparent $\theta$ of ~21.5° attained. Since both hydrated salt cation and nanoparticles were in the displacing fluid, the dynamic improvement was likely a co-contribution from both chemicals interacting at the water wedge.

It is important to note that the crude oil droplet remained almost spherical shape at the final stage of the dewetting dynamics, see **Fig. 6**, reflecting no domination of other forces (e.g., strong capillary and viscous) in action, as observed in other works [2,38]. Without detectable residual oil left on the substrate, the detaching droplet floated up gently. Given an interfacially active property of the pNIPAM nanoparticles, the $\sigma$ was reduced by almost two-fold (21.7 → 12.3 mN/m) when compared to the brine-only system, see Tangparitkul et al. [10] for 2000 ppm NaCl brine-only. The calculated substrate-oil adhesion force was not decreased by a substantial degree (19.0 → 15.0 µN) as the $\sigma$ reduction did. The calculated adhesion force in the current fluid (i.e., the 0.0005 wt.% pNIPAM + SDS blend in the 2000 ppm NaCl brine) is not even lower than the droplet buoyancy (3.8 µN), confirming a negligible buoyant-gravitational control but rather the non-DLVO force induced displacement of crude oil droplet per se – a benefit of using interfacially active nanoparticles, with potential coupling effect in the low-salinity brine.

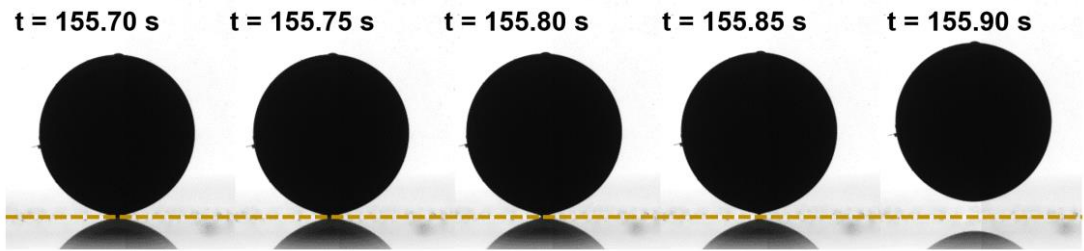

**Figure 6.** Development of the detachment of a crude oil droplet in the presence of the pNIPAM + SDS blend in 2000 ppm NaCl brine at the final stage of dewetting dynamics, as observed by an optical tensiometer. The oil droplet receded gradually while maintaining an axisymmetric spherical shape. The brown dashed line indicates the solid-liquid interface.

When further considering the same monovalent brine-only system in the previous study [10], the initial receding rate was 9.46 °/s and the steady-state $\theta$ was 18.1°; both are comparable to those of the current studied fluid – the pNIPAM + SDS blend in the same monovalent brine. This implies that the initial period of the displacement in the current blend was dominantly controlled by the brine species (i.e., due to the hydration force) rather than the nanoparticles, while in the later period when the oil droplet about to detach was not since the brine-only fluid has found no detachment [10]. As such, the nanoparticle-induced disjoining pressure is thought to actively govern in the later period only, which gradually invades the water wedge and causes detachment.

## 4. Conclusions

The crude oil displacement dynamics of crude oil droplet dynamic dewetting on flat glass substrate were investigated for the influences of two species of EOR chemicals, namely nanoparticles and low-salinity brine. Based on previous work, the mixture of the two was the focus of the current study, where the interfacial phenomena were elucidated and the influences of the chemical-induced colloidal forces on the oil displacement dynamics were considered. Based on the observations and findings in this study, the following conclusions can be drawn:

1. The pNIPAM + SDS blend dispersed in Milli-Q water at a low component concentration has a greater oil displacement due to a contribution from the nanoparticle-induced structural disjoining pressure. At a higher concentration, the nanoparticles preferred to partition at the oil-water interface, leading to the demoting effect of polymer bridging to the solid substrate.

2. When dispersed in the low-salinity NaCl brine, the nanoparticles did not aggregate, and the co-existence of the bine and the nanoparticles resulted in greater oil dewetting with the eventual detachment of the crude oil droplet. When dispersed in the divalent brine counterpart, much lower performance was observed on the oil displacement as a result of stronger bridging forces between the oil-water and water-solid interfaces promoted by the divalent salt cations.

3. Considering the coupling effect of the nanoparticles in the NaCl brine, the enhancing mechanisms of oil displacement are thought to comprise two stages. At the initial period of dewetting, the monovalent salt hydration force controlled the crude oil droplet dewetting, while at the later period – with a shaper water wedge – the nanoparticle-induced structural force caused the detachment of the oil droplet further from the substrate – without pinching the droplet.

Overall, the current study highlights the coupling effect of interfacially active nanoparticles and low-salinity brine on the dewetting of heavy crude oil based on a series of experimental observations. Adding the right amount of nanoparticles to an optimal brine could be an option for faster and greater fluid displacement. One practical application of the investigated chemicals is the use of nanofluids concurrently with low-salinity fluids for the EOR purpose. Furthermore, the findings are not limited to only EOR application but also some others, such as detergency and other forms of geological storage.

**Acknowledgements**

Financial support for this work is greatly acknowledged with contribution from Chiang Mai University through Chiang Mai Research Center for Carbon Capture and Storage (Chiang Mai CCS) (No. RG 58/2566). Dr. Mohammed Jeraal (University of Cambridge) and Dr. Chris S. Hodges (University of Leeds) are thanked for their assistances on pNIPAM nanoparticle preparation and QCM experiment, respectively.